\documentclass[graybox]{svmult}

\usepackage{mathptmx}       
\usepackage{helvet}         
\usepackage{courier}        
\usepackage{type1cm}        

\usepackage{makeidx}         
\usepackage{graphicx}        
\usepackage{multicol}        
\usepackage[bottom]{footmisc}

\makeindex            

\begin{document}

\title*{Perturbations of Kantowski-Sachs models with a cosmological constant}
\author{Z. 
Keresztes, M. 
Forsberg, M. 
Bradley, P. 
K.S. Dunsby and L. 
\'{A}. Gergely}
\institute{Zolt\'{a}n Keresztes, L\'{a}szl\'{o} \'{A}. Gergely \at Department of Theoretical Physics
and Department of Experimental Physics, University of Szeged, Szeged, Hungary \email{zkeresztes@titan.physx.u-szeged.hu, gergely@physx.u-szeged.hu}
\and Mats Forsberg, Michael Bradley \at Department of Physics, Ume\aa\ University, Sweden 
\email{mats.forsberg@physics.umu.se, michael.bradley@physics.umu.se}
\and Peter K.S. Dunsby \at Department of Mathematics and Applied Mathematics, University of Cape
Town, South Africa \email{peter.dunsby@uct.ac.za}
}

\maketitle

\abstract{We investigate perturbations of Kantowski-Sachs 
models with a positive cosmological constant, 
using the gauge invariant 1+3 and 1+1+2 covariant splits of spacetime 
together with a harmonic decomposition. The perturbations are assumed to be 
vorticity-free and of perfect fluid type, but otherwise include general scalar, vector 
and tensor modes. In this case the set of equations can be reduced to six evolution 
equations for six harmonic coefficients. }

\section{Introduction}

In this work we consider perturbations of Kantowski-Sachs 
models with a positive cosmological constant. Some of these models can 
undergo an anisotropic bounce where the universe changes from a contracting to 
an expanding phase. A simple argument used by B\"orner and Ehlers, \cite{BornerEhlers}, 
to show that an isotropic bouncing universe is excluded by observations does not hold for the 
Kantowski-Sachs models \cite{KASAscalar}. Hence it is of interest to study the evolution and propagation 
of perturbations in these models and their possible effects on observables, like the Sachs-Wolfe effect \cite{SachsWolfe}.
To do this we use the 1+3 and 1+1+2 covariant splits of spacetime,
\cite{EllisBruni, EllisvanElst,ClarksonBarrett, Clarkson}, that are suitable for perturbation theory,
as they employ variables that vanish on the background and hence their 
perturbations are gauge invariant \cite{StewartWalker}.  The perturbations are assumed to be 
vorticity-free and of perfect fluid type, but otherwise include general scalar, vector 
and tensor modes. The evolution equations for the perturbative variables are then derived
in terms of harmonics.

\section{The 1+3 and 1+1+2 covariant formalisms}

A covariant formalism for the 1+3 split of spacetimes with a preferred timelike
vector, $u^a$, was developed in \cite{EllisBruni, EllisvanElst}. The projection operator onto the perpendicular 3-space is given by
$h_a^b=g_a^b+u_au^b$ . With the help of this vectors and tensors can be covariantly decomposed into
"spatial" and "timelike" parts.
The covariant time derivative and projected spatial derivative are given by
\begin{eqnarray}
\dot \psi_{a..b}\equiv u^c\nabla_c\psi_{a...b}
\quad \hbox{and} \quad
D_c\psi_{a...b}\equiv h^f_c h^d_a...h^e_b\nabla_f\psi_{d...e}
\end{eqnarray}
respectively.
The covariant derivative of the 4-velocity, $u^a$, can be decomposed as
\begin{equation}
\nabla_a u_b=-u_a A_b+D_a u_b=-u_a A_b+\frac{1}{3}\theta h_{ab}+\omega_{ab}+\sigma_{ab}
\end{equation}
where the kinematic quantities of $u^a$, acceleration, expansion,vorticity and shear are defined by $A_a\equiv u^b\nabla_b u_a$, $\theta \equiv D_a u^a$, $\omega_{ab}\equiv D_{[a}u_{b]}$, and $\sigma_{ab}\equiv D_{<a}u_{b>}$ respectively. These quantities, together with the Ricci tensor 
(expressed via the Einstein equations by 
energy density $\mu$ and pressure $p$ for a perfect fluid) and the electric, 
$E_{ab}\equiv  C_{acbd}u^c u^d$, and magnetic, $H_{ab}\equiv \frac{1}{2}\eta_{ade}C^{de}\!\!_{bc}u^c$,
parts of the Weyl tensor, are then used as dependent variables.
From the Ricci and Bianchi identities one obtains evolution equations in the $u^a$ direction and constraints.

A formalism for a further split (1+2) with respect to a spatial vector $n^a$ (with $u^a n_a=0$) was
developed in \cite{ClarksonBarrett, Clarkson}. Projections perpendicular to $n^a$ are made with
$N_a^b=h_a^b-n_an^b$, and in an analogous way to above "spatial" vectors and tensors may
be decomposed into scalars along $n^a$ and perpendicular two-vectors and symmetric, trace-free two-tensors as
$A^a={\cal A}n^a+{\cal A}^a$ , $\omega^a=\Omega n^a+\Omega^a$,
$\sigma_{ab}=\Sigma(n^a n^b-\frac{1}{2}N_{ab})+2\Sigma_{(a}n_{b)}+\Sigma_{ab}$		
and similarly for $E_{ab}$ and $H_{ab}$ in terms of ${\cal E}$, ${\cal E}_a$, ${\cal E}_{ab}$
and ${\cal H}$, ${\cal H}_a$, ${\cal H}_{ab}$ respectively.
Derivatives along and perpendicular to $n^a$ are
\begin{eqnarray}
\hat\psi_{a...b}\equiv n^c D_c\psi_{a...b}=n^ch^f_c h^d_a...h^e_b\nabla_f\psi_{d...e}
\;\; \hbox{and} \;\;
\delta_c\psi_{a...b}\equiv N_c^f N_a^d...N_b^e D_f\psi_{d...e}
\end{eqnarray}
respectively. Similarly to the decomposition of $\nabla_a u_b$, $D_a n_b$ and $\dot n_a$
can be decomposed into further `kinematical' quantities of $n^a$ as
\begin{equation}
		D_a n_b =n_a a_b+\frac{1}{2}\phi N_{ab}+\xi\epsilon_{ab}+\zeta_{ab} \quad 
\hbox{and} \quad \dot n_a={\cal A}u_a+\alpha_a
		\end{equation}
where $a_a\equiv \hat n_a$, $\phi \equiv \delta_a n^a$, $\xi\equiv\frac{1}{2}\epsilon^{cabd}\delta_a n_bu_cn_d$, $\zeta_{ab}\equiv \delta_{\{a}n_{b\}}$, ${\cal A}\equiv n^aA_a$, $\alpha_a\equiv N_a^b\dot n_b$.

The Ricci and Bianchi identities are then written as
evolution and propagation equations in the $u^a$ and $n^a$ directions and constraints.

\section{Perturbations of Kantowski-Sachs}
As backgrounds we take the Locally Rotationally Symmetric (LRS) 
Kantowski-Sachs cosmologies \cite{KantowskiSachs}
\begin{equation}
ds^2 = -dt^2 + a_1^2(t)dz^2+a_2^2(t)\left(d\vartheta^2+\sin^2\theta d\varphi^2\right) \, 
\end{equation}
with cosmological constant $\Lambda>0$ and matter given by a perfect fluid with barytropic equation $p=p(\mu)$. 
The shear $\Sigma$, energy density $\mu$ and the expansion $\theta$ 
evolve as
\begin{eqnarray}\label{eq01}
\dot \Sigma =-\frac{1}{2}\Sigma^2-\frac{2}{3}\Sigma\theta-{\cal E},
 \;\;
\dot \mu=-\theta (\mu + p), \;\;
\dot \theta =(\Lambda-\frac{1}{2}\mu-\frac{3}{2}p) -\frac{1}{3}\theta^2-\frac{3}{2}\Sigma^2 
\end{eqnarray}
where the electric part of the Weyl tensor is 
${\cal E}=- \frac{2}{3}\mu- \frac{2}{3}\Lambda-\Sigma^2+\frac{2}{9}\theta^2+\frac{1}{3}\Sigma\theta$.

Instead of the background variables $\theta, \Sigma, {\cal E}, \mu$ we 
use their gradients
\begin{equation}
		W_a\equiv\delta_a\theta, \quad V_a\equiv\delta_a\Sigma, \quad
		X_a\equiv\delta_a{\cal E}, \quad \mu_a\equiv\delta_a\mu \, ,
		\end{equation}
which vanish on the background and hence are gauge 
invariant
(the derivatives $\hat \theta\equiv n^a D_a\theta$ etc. can be given in terms of the
$\delta_a$ derivatives due to commutation relations in the case of no vorticity).
Similar variables vanishing on the background are
		$a_a, \phi, \xi, \zeta_{ab}, \alpha_a, {\cal A}, {\cal A}_a$,	
		$\Sigma_a, \Sigma_{ab}, {\cal E}_a, {\cal E}_{ab}, {\cal H}, {\cal H}_a, {\cal H}_{ab}$
where $a_a$ can be put to zero by choice of frame.

The scalar, vector and tensor variables are expanded in harmonics according to
\begin{eqnarray}\nonumber
\Psi&=&\sum\limits_{k_{\parallel}, k_{\perp}}\Psi_{k_{\parallel} k_{\perp}}P_{k_\parallel}Q_{k_\perp}
\, , \quad
\Psi_a=\sum\limits_{k_{\parallel}, k_{\perp}}P_{k_\parallel}
\left(\Psi^V_{k_{\parallel} k_{\perp}}Q^{k_\perp}_a+\overline\Psi^V_{k_{\parallel} k_{\perp}}
\overline Q^{k_\perp}_a\right) \, , \\
\Psi _{ab}&=&\sum\limits_{k_{\parallel },k_{\perp }}P_{k_{\parallel }}\
\left( \Psi _{k_{\parallel },k_{\perp }}^{T}Q_{ab}^{k_{\perp }}+\overline{%
\Psi }_{k_{\parallel },k_{\perp }}^{T}\overline{Q}_{ab}^{k_{\perp }}\right)
\end{eqnarray}
where $Q_{k_{\perp}}$, $Q_{a}^{k_{\perp }}$, $\overline{Q}_{a}^{k_{\perp }}$,
 $Q_{ab}^{k_{\perp }}$ and $\overline{Q}_{ab}^{k_{\perp }}$
are harmonics on the 2-spheres of constant $z$ and $P_{k_{\parallel }}$ the corresponding
expansion functions in the $z$-direction.

All coefficients can be given in terms of
$\mu _{k_{\parallel },k_{\perp }}^{V}$, $\Sigma _{k_{\parallel
},k_{\perp }}^{T}$, $\mathcal{E}_{k_{\parallel },k_{\perp }}^{T}$, 
$\overline{\mathcal{H}}_{k_{\parallel},k_{\perp }}^{T}$ and
$\overline{\mathcal{E}}_{k_{\parallel },k_{\perp }}^{T}$, $\mathcal{H}%
_{k_{\parallel },k_{\perp }}^{T}$, so the system has six degrees of freedom. 
The first four coefficients form a closed system of evolution equations
coupled to the density gradient, in agreement with the results for scalar perturbations in
\cite{KASAscalar}. This reads
\begin{eqnarray}\nonumber
\dot{\mu}_{k_{\parallel },k_{\perp }}^{V}\!\! &=&\left[ \frac{\Sigma }{2}%
\left( 1-6\frac{\mu +p}{B}\right) \!-\frac{4\theta }{3}\right] \mu
_{k_{\parallel },k_{\perp }}^{V}\!+
\\\nonumber
&&\frac{a_{2}}{2}\left( \mu +p\right) 
\left[ \left( 1-C\right) \!\left( B\Sigma
_{k_{\parallel },k_{\perp }}^{T}\!+\mathcal{E}_{k_{\parallel },k_{\perp
}}^{T}\right) -P\overline{\mathcal{H}}_{k_{\parallel },k_{\perp }}^{T}\right]
,  \label{1}
\\\nonumber
\dot{\Sigma}_{k_{\parallel },k_{\perp }}^{T}& =&-\frac{1}{%
a_{2}\left( \mu +p\right) }\frac{dp}{d\mu}\mu _{k_{\parallel },k_{\perp }}^{V}  
+\!\left( \Sigma -\frac{2\theta }{3}\right) \Sigma _{k_{\parallel
},k_{\perp }}^{T}-\mathcal{E}_{k_{\parallel },k_{\perp }}^{T}\ ,
\\\nonumber
\dot{\mathcal{E}}_{k_{\parallel },k_{\perp }}^{T}\! &=&\!-\frac{3\Sigma }{%
2a_{2}B}\mu _{k_{\parallel },k_{\perp }}^{V}-\frac{\mu +p}{2}\Sigma
_{k_{\parallel },k_{\perp }}^{T}  
-\frac{3}{2}\left( F+\Sigma C\right) \mathcal{E}_{k_{\parallel },k_{\perp
}}^{T}+\frac{P}{2}\overline{\mathcal{H}}_{k_{\parallel },k_{\perp }}^{T}\ ,
\\
\dot{\overline{\mathcal{H}}}_{k_{\parallel },k_{\perp }}^{T}\!\!\!\! &=&-%
\frac{ik_{\parallel }}{a_{1}a_{2}B}\mu _{k_{\parallel },k_{\perp }}^{V}-R%
\overline{\mathcal{H}}_{k_{\parallel },k_{\perp }}^{T}  
-\frac{ik_{\parallel }}{a_{1}}\left[ 1-\frac{3}{2}\left( C-\frac{\mathcal{E%
}}{B}\right) \right] \mathcal{E}_{k_{\parallel },k_{\perp }}^{T}\ ,
\label{6}
\end{eqnarray}
where we have introduced the notations
$B=\frac{2k_{\parallel }^{2}}{a_{1}^{2}}+\frac{k_{\perp }^{2}}{a_{2}^{2}}+
\frac{9}{2}\Sigma ^{2}+3\mathcal{E}$,
$\quad C=B^{-1}\left( \frac{2-k_{\perp }^{2}}{a_{2}^{2}}+3\mathcal{E}\right)$,
$D=C+\frac{\mu +p}{B}$,
$E=\frac{\Sigma }{2}\left( C-\frac{\mathcal{E}}{B}\right) +\frac{\theta 
\mathcal{E}}{3B}$,
$F=\Sigma +\frac{2\theta }{3}$,
$P=\frac{a_{1}}{ik_{\parallel }}\left[ \frac{2k_{\parallel }^{2}}{a_{1}^{2}}%
\left( 1-C\right) -\frac{k_{\perp }^{2}}{a_{2}^{2}}\frac{2-k_{\perp }^{2}}{%
a_{2}^{2}B}\right]$ and
$R =\frac{3}{2}F-\left( \Sigma +\frac{\theta }{3}\right) \frac{k_{\perp
}^{2}}{a_{2}^{2}B} 
-\frac{1}{2B}\left( \Sigma -\frac{2\theta }{3}\right) \left( D-\frac{%
2k_{\parallel }^{2}}{a_{1}^{2}}\right)$.
The two last coefficents form a closed system for free waves
\begin{eqnarray}\nonumber
\dot{\overline{\mathcal{E}}}_{k_{\parallel },k_{\perp }}^{T}\!\!\!&=&-\frac{3}{%
2}\!\left( F\!+\!\Sigma D\!\right) \!\overline{\mathcal{E}}_{k_{\parallel
},k_{\perp }}^{T}\!\!\!+\frac{ik_{\parallel }}{a_{1}}\left( 1-D\right) 
\mathcal{H}_{k_{\parallel },k_{\perp }}^{T},  \label{4} \\
\dot{\mathcal{H}}_{k_{\parallel },k_{\perp }}^{T} &=&-\frac{a_{1}}{%
2ik_{\parallel }}\left( \frac{2k_{\parallel }^{2}}{a_{1}^{2}}-BC+9\Sigma
E\right) \overline{\mathcal{E}}_{k_{\parallel },k_{\perp }}^{T}  
-\frac{3}{2}\left( 2E+F\right) \mathcal{H}_{k_{\parallel },k_{\perp
}}^{T}\ .  \label{5}
\end{eqnarray}%

These sets of equations can be used to study the propagation of gravitational waves
and the coupling between scalar and tensor perturbations.
Furthermore, from the null geodesics of photons, equations for the redshift in different directions
can be given  completely in terms of the 1+1+2 quantities. From their solutions the Sachs-Wolfe
effect and the corresponding variations in the CMB temperature can be calculated.

\begin{acknowledgement}
ZK was supported by OTKA grant no. 100216
\end{acknowledgement}

\end{document}